\documentclass[prd,nofootinbib,twocolumn,showpacs,preprintnumbers,floatfix,amsmath,amssymb,amsfonts]{revtex4}

\usepackage{graphicx}
\usepackage{amsmath}
\usepackage{dcolumn}
\usepackage{epsfig}
\usepackage{psfrag}
\usepackage{lineno}
\usepackage{multirow}
\RequirePackage{xspace}

\DeclareGraphicsExtensions{.epsi,.eps,.ps,.eps.gz,.ps.gz,.gif,.epsi.gz}
\def\vckm       {\ensuremath{ {V}_{CKM}}}

\def\theckmmatrix  {\ensuremath{ \left( \begin{array}{ccc} \vud & \vus & \vub \\ \vcd & \vcs & \vcb \\ \vtd & \vts & \vtb \end{array}\right).}}

\def\CPV {\ensuremath{CPV}}

\RequirePackage{xspace}
\usepackage{color}
\definecolor{Red}{rgb}{1,0,0}
\definecolor{Magenta}{rgb}{0.5,0,0.5}
\definecolor{Blue}{rgb}{0,0,1}
\definecolor{Green}{rgb}{0,1,0}

\usepackage{relsize}
\def\babar{\mbox{\slshape B\kern-0.1em{\smaller A}\kern-0.1em
    B\kern-0.1em{\smaller A\kern-0.2em R}}\xspace}

\def\lhcb {LHCb\xspace}

\def\Kbar  {\kern 0.2em\overline{\kern -0.2em K}{}\xspace}

\def\Kz    {\ensuremath{K^0}\xspace}
\def\Kzb   {\ensuremath{\Kbar^0}\xspace}
\def\KzKzb {\ensuremath{\Kz \kern -0.16em \Kzb}\xspace}
\def\Kp    {\ensuremath{K^+}\xspace}
\def\Km    {\ensuremath{K^-}\xspace}

\def\KpKm  {\ensuremath{\Kp \kern -0.16em \Km}\xspace}

\def\Dbar    {\kern 0.2em\overline{\kern -0.2em D}{}\xspace}

\def\Dz      {\ensuremath{D^0}\xspace}
\def\Dzb     {\ensuremath{\Dbar^0}\xspace}
\def\DzDzb   {\ensuremath{\Dz {\kern -0.16em \Dzb}}\xspace}
\def\Dp      {\ensuremath{D^+}\xspace}
\def\Dm      {\ensuremath{D^-}\xspace}

\def\DpDm    {\ensuremath{\Dp {\kern -0.16em \Dm}}\xspace}

\def\Bbar    {\kern 0.18em\overline{\kern -0.18em B}{}\xspace}

\def\Bz      {\ensuremath{B^0}\xspace}
\def\Bzb     {\ensuremath{\Bbar^0}\xspace}
\def\BzBzb   {\ensuremath{\Bz {\kern -0.16em \Bzb}}\xspace}
\def\Bu      {\ensuremath{B^+}\xspace}
\def\Bub     {\ensuremath{B^-}\xspace}

\def\BpBm    {\ensuremath{\Bu {\kern -0.16em \Bub}}\xspace}
\def\Bs      {\ensuremath{B_s}\xspace}
\def\Bsb     {\ensuremath{\Bbar_s}\xspace}
\def\BsBsb   {\ensuremath{\Bs {\kern -0.16em \Bsb}}\xspace}

\def\BorBbar    {\kern 0.18em\optbar{\kern -0.18em B}{}\xspace}
\def\DorDbar    {\kern 0.18em\optbar{\kern -0.18em D}{}\xspace}
\def\KorKbar    {\kern 0.18em\optbar{\kern -0.18em K}{}\xspace}

\mathchardef\Upsilon="7107

\def\FourS {\ensuremath{\Upsilon{(4S)}}\xspace}

\mathchardef\Deltares="7101
\mathchardef\Xi="7104
\mathchardef\Lambda="7103
\mathchardef\Sigma="7106
\mathchardef\Omega="710A

\def\Deltabar{\kern 0.25em\overline{\kern -0.25em \Deltares}{}\xspace}
\def\Lbar{\kern 0.2em\overline{\kern -0.2em\Lambda\kern 0.05em}\kern-0.05em{}\xspace}
\def\Sigbar{\kern 0.2em\overline{\kern -0.2em \Sigma}{}\xspace}
\def\Xibar{\kern 0.2em\overline{\kern -0.2em \Xi}{}\xspace}
\def\Obar{\kern 0.2em\overline{\kern -0.2em \Omega}{}\xspace}
\def\Nbar{\kern 0.2em\overline{\kern -0.2em N}{}\xspace}
\def\Xb{\kern 0.2em\overline{\kern -0.2em X}{}\xspace}

\iffalse

\else

\fi

\newcommand{\tev}{\ensuremath{\mathrm{\,Te\kern -0.1em V}}\xspace}
\newcommand{\gev}{\ensuremath{\mathrm{\,Ge\kern -0.1em V}}\xspace}
\newcommand{\mev}{\ensuremath{\mathrm{\,Me\kern -0.1em V}}\xspace}
\newcommand{\kev}{\ensuremath{\mathrm{\,ke\kern -0.1em V}}\xspace}
\newcommand{\ev}{\ensuremath{\mathrm{\,e\kern -0.1em V}}\xspace}
\newcommand{\gevc}{\ensuremath{{\mathrm{\,Ge\kern -0.1em V\!/}c}}\xspace}
\newcommand{\mevc}{\ensuremath{{\mathrm{\,Me\kern -0.1em V\!/}c}}\xspace}
\newcommand{\gevcc}{\ensuremath{{\mathrm{\,Ge\kern -0.1em V\!/}c^2}}\xspace}
\newcommand{\mevcc}{\ensuremath{{\mathrm{\,Me\kern -0.1em V\!/}c^2}}\xspace}

\def\invfb   {\ensuremath{\mbox{\,fb}^{-1}}\xspace}
\def\invab   {\ensuremath{\mbox{\,ab}^{-1}}\xspace}

\def\mus  {\ensuremath{\rm \,\mus}\xspace}

\def\mus        {\ensuremath{\,\mu{\rm s}}\xspace}    

\def\to                 {\ensuremath{\rightarrow}\xspace}

\def\pep2{PEP-II}

\def\gsim{{~\raise.15em\hbox{$>$}\kern-.85em
          \lower.35em\hbox{$\sim$}~}\xspace}
\def\lsim{{~\raise.15em\hbox{$<$}\kern-.85em
          \lower.35em\hbox{$\sim$}~}\xspace}

\newcommand\vud {\ensuremath{V_{ud}}}
\newcommand\vus {\ensuremath{V_{us}}}
\newcommand\vub {\ensuremath{V_{ub}}}
\newcommand\vcd {\ensuremath{V_{cd}}}
\newcommand\vcs {\ensuremath{V_{cs}}}
\newcommand\vcb {\ensuremath{V_{cb}}}
\newcommand\vtd {\ensuremath{V_{td}}}
\newcommand\vts {\ensuremath{V_{ts}}}
\newcommand\vtb {\ensuremath{V_{tb}}}
\def\vckm       {\ensuremath{ {V}_{CKM}}}

\def\theckmmatrix  {\ensuremath{ \left( \begin{array}{ccc} \vud & \vus & \vub \\ \vcd & \vcs & \vcb \\ \vtd & \vts & \vtb \end{array}\right)}}

\xspace

\def\jetset74   {\mbox{\tt Jetset \hspace{-0.5em}7.\hspace{-0.2em}4}\xspace}

\long\def\inst#1{\par\nobreak\kern 4pt\nobreak
    {\it #1}\par\vskip 10pt plus 3pt minus 3pt}

\begin{document}


{\pagestyle{empty}

\par\vskip 3cm

\title{
\Large \boldmath
The Time-Dependent \boldmath{$CP$} Violation in Charm
}

\author{G. Inguglia}
\affiliation{Queen Mary, University of London, Mile End Road, E1 4NS, United Kingdom}

\date{\today}

\begin{abstract}
%
%
A model which describes the time-dependent $CP$ formalism in $D^0$ decays has recently been proposed. There it has been highlighted a possible measurement of the angle $\beta_c$, in the charm unitarity triangle, using the decays $D^0\to K^+ K^-$ and $D^0\to \pi^+ \pi^-$, and a measurement of the mixing phase $\phi_{MIX}$. The same method can be used to measure the value of the parameter $x$, one of the two parameters defining charm mixing. We numerically evaluate the impact of a time-dependent analysis in terms of the possible outcomes from present and future experiments. We consider the scenarios of correlated $D^0$ mesons production at the center of mass energy of the $\Psi(3770)$ at Super$B$, uncorrelated production at the center of mass energy of the $\Upsilon(4S)$ at Super$B$ and Belle II, and LHCb. Recently a hint of direct $CP$ violation in charm decays was reported by the LHCb collaboration, we estimate the rate of time-dependent asymmetry that could be achieved using their available data, and we generalise the result for the full LHCb program. We conclude that LHCb is already able to perform a first measurement of $\beta_{c,eff}$, and slightly improve the present constraints on the parameters $x$ and $\phi_{MIX}$. A more precise determination of $\beta_{c,eff}$, $\phi_{MIX}$ and $x$ will require a larger data sample, and most probably the cleaner environment of the new high luminosities $B$-factories (both Super$B$ and Belle II) will be needed. We show that Super$B$ will be able to measure $\beta_{c,eff}$ and $\phi_{MIX}$ with a precision of $1.4^o$ and improve the precision on $x$ by a factor of two.
\end{abstract}


\maketitle

\vfill

}

\setcounter{footnote}{0}

\section{Introduction}
\label{sec:intro}

Since the discovery in 1964 of $CP$ violation in the Kaon system~\cite{Fitch}, $CP$ violation has been observed also in the $B$ meson system~\cite{CPB}~\cite{CPBe}. In the charm sector, $CP$ violation has long been expected to be too small to be observed at precision available until recently when, the LHCb collaboration has reported a difference in direct $CP$ asymmetries in $D^0\to K^+ K^-$ and $D^0\to \pi^+ \pi^-$ that is $3.5 \sigma$ from the $CP$ conserving hypothesis~\cite{note}.  
In~\cite{bim} the standard model (SM) description of these decays using the same CKM paradigm that provides a rather satisfactory description of such decays of $\Bz$ mesons is considered. Since the LHCb result, a broader view of this paradigm that might accomodate the large asymmetry is examined in~\cite{kagan}. It is clear that, in order to understand the nature of \CPV in \Dz decays, measurements of weak phases in these decays are essential. In~\cite{bim}, it is proposed that, as with \Bz decays, time-dependent CP asymmetries in \Dz decays may provide the most direct way to measure these phases. In this paper, we further examine the precision that might be anticipated in four experimental scenarios that are likely to be available over the coming decade to evaluate the rate of time-dependent $CP$ asymmetries in $D^0\to K^+ K^-$ and $D^0\to \pi^+ \pi^-$ in the three proposed environments (Super$B$, LHCb, Belle II).

\section{Time-dependent \boldmath{$CP$} violation in the charm sector}
\label{sec:tdcpc}
In the standard model (SM), $CP$ violation is described in terms of the complex phase appearing in Cabibbo-Kobayashi-Maskawa (CKM) matrix~\cite{Cabibbo}~\cite{Kob-Mask}. The matrix is a unitary $3\times3$ matrix which provides a description of quark mixing in terms of the coupling strengths for up-to-down quark type transitions, and it may be written as

\begin{eqnarray}
\vckm = \theckmmatrix.\label{eq:ckmmatrix}
\end{eqnarray}
Within this framework the probability to observe a transition between a quark $q$ to a quark $q'$ is proportional to $|V_{qq'}|^2$.

\subsection{Buras parametrisations of the CKM Matrix}
In Ref.~\cite{bim} the CKM matrix has been written using the $"Buras"$ parametrisation~\cite{Buras}:
\begin{widetext}
\begin{eqnarray}
\vckm =  \ensuremath\left (
 \begin{array}{ccc }
 1 - \lambda^2 / 2 -\lambda^4/8     &   \lambda    & A\lambda^3 (\bar\rho - i\bar\eta)+A\lambda^5(\bar\rho - i\bar\eta)/2 \\
 -\lambda +A^2\lambda^5[1 - 2(\bar\rho+i\bar\eta)]/2 & 1 - \lambda^2/ 2 - \lambda^4(1+4A^2)/8 & A \lambda^2  \\
  A\lambda^3[1-(\bar\rho +i\bar\eta)]  & -A \lambda^2 + A\lambda^4[1-2(\bar\rho+i\bar\eta)]/2   & 1 -A^2\lambda^4/2
 \end{array}
\right) + {\cal O}(\lambda^6),
\label{eq:ckmmatrixburasrhobaretabar}
\end{eqnarray}
\end{widetext}

We adopt the convention of writing the CKM matrix in terms of $\overline{\rho}$ and $\overline{\eta}$ because these represent the coordinates of the apex of the well known $bd$ unitarity triangle. Since unitarity triangles are mathematically exact, it is very important to measure their angles and sides to verify unitarity. One of the six unitarity relationships of the CKM matrix may be written as
\begin{eqnarray}
V_{ud}^* V_{cd} + V_{us}^* V_{cs} + V_{ub}^* V_{cb} = 0, \label{eq:charmtriangle}  
\end{eqnarray}
where Eq.~(\ref{eq:charmtriangle}) represents the \textit{cu} triangle that we will call the \textit{charm} unitarity triangle or simply \textit{charm} triangle. The internal angles of this triangle are given by
\begin{eqnarray}
\alpha_c &=& \arg\left[- V_{ub}^*V_{cb}/ V_{us}^*V_{cs} \right], \label{eq:alphac}\\
\beta_c  &=& \arg\left[- V_{ud}^*V_{cd} / V_{us}^*V_{cs} \right], \label{eq:betac}\\
\gamma_c &=& \arg\left[- V_{ub}^*V_{cb} / V_{ud}^*V_{cd}\right]. \label{eq:gammac}
\end{eqnarray}
In Ref.~\cite{bim} we proposed the measurement of $\beta_{c,eff}$ using time-dependent $CP$ asymmetries in charm decays and using the results of Global CKM fits, predicted that:
\begin{eqnarray}
    \beta_c &=& (0.0350\pm 0.0001)^{\circ}. 
\end{eqnarray}
On comparing Eq.~(\ref{eq:betac}) with Eq.~(\ref{eq:ckmmatrixburasrhobaretabar}), one can 
see that $V_{cd} = V_{cd} e^{i(\beta_c-\pi)}$ in this convention.
\subsection{Time-dependent formalism}
We consider two different cases of $D^0$ meson production: un-correlated and correlated $D^0$ production. Un-correlated $D^0$'s are produced from the decays of $B$ mesons in electron-positron colliders when particles are collided at a center of mass energy corresponding to the $\Upsilon(4S)$ resonance, from $c\overline{c}$ continuum, or in hadrons collider. The correlated $D^0$ mesons are produced in an electron-positron machine running at a center of mass energy corresponding to the $\Psi(3770)$ resonance. The time evolution for both situations is given by~\cite{bim}

\begin{widetext}
\textit{un-correlated case}
\begin{eqnarray}
\Gamma_{\pm} \propto e^{-\Gamma_1 t}\left[ \frac{\left(1+e^{\Delta\Gamma t} \right)}{2} + \frac{Re(\lambda_{f})}{1+|\lambda_{f}|^2}\left(1-e^{\Delta\Gamma t} \right) \pm e^{\Delta\Gamma t / 2}\left(\frac{1-|\lambda_{f}|^2}{1+|\lambda_{f}|^2}\cos \Delta M t - \frac{2 Im(\lambda_{f})}{1+|\lambda_{f}|^2}\sin \Delta M t  \right) \right],  \label{EQ:p0toffinal}
\end{eqnarray}
\end{widetext}

\begin{widetext}
\textit{correlated case}
\begin{eqnarray}
\Gamma_{\pm} &\propto& e^{-\Gamma_1 |\Delta t|}\left[ \frac{h_+}{2} + \frac{ Re(\lambda_{f})}{1+|\lambda_{f}|^2}h_- \pm e^{\Delta\Gamma \Delta t / 2}\left(\frac{1-|\lambda_{f}|^2}{1+|\lambda_{f}|^2}\cos \Delta M \Delta t - \frac{2 Im(\lambda_{f})}{1+|\lambda_{f}|^2}\sin \Delta M \Delta t  \right) \right]. \label{EQ:p0toffinaldeltat}
\end{eqnarray}
\end{widetext}

\begin{figure}[!ht]
\begin{center}
\resizebox{9.cm}{!}{
\includegraphics{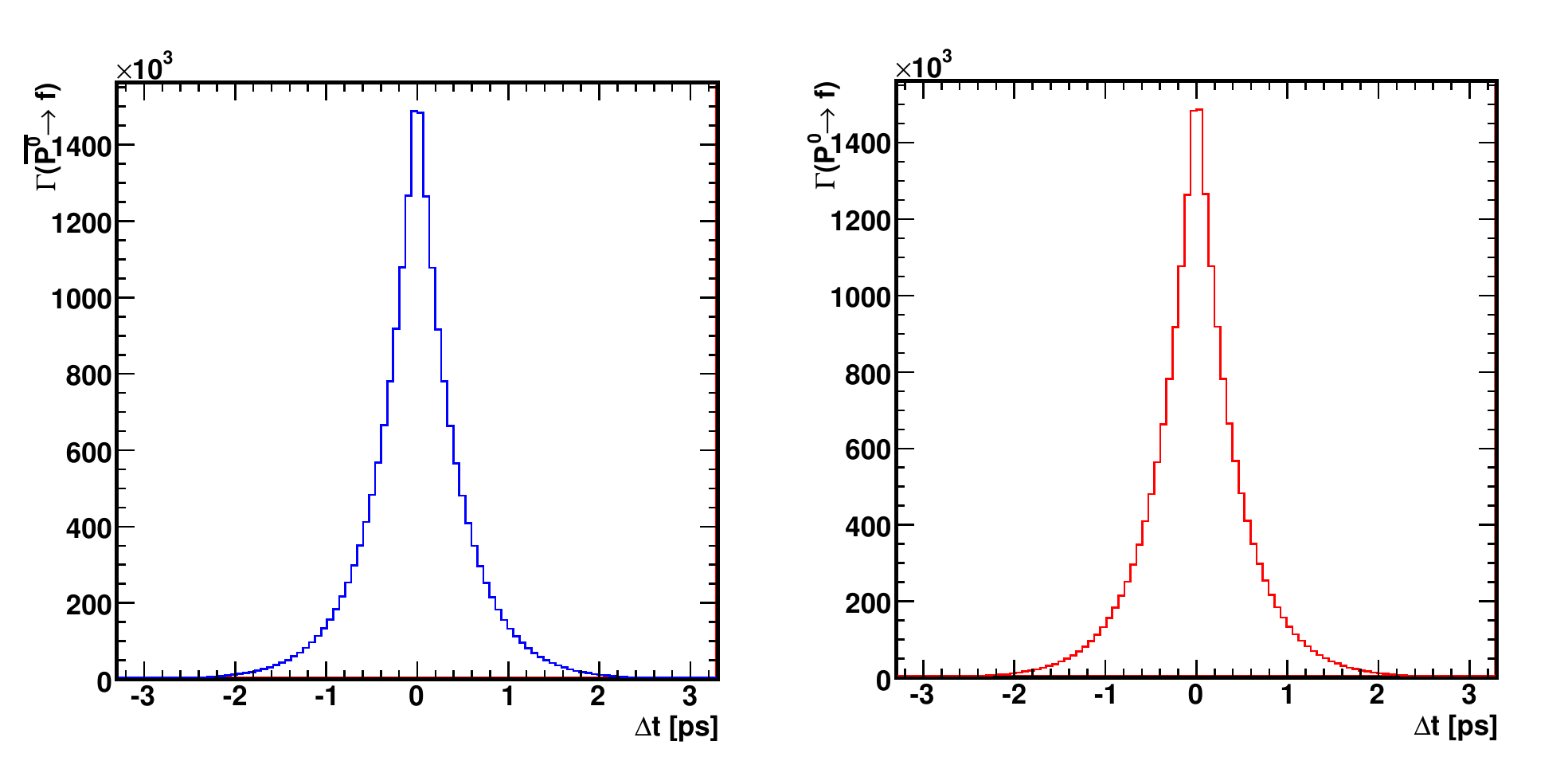}
}
\caption{Generated distributions according to our formulas for $\overline{D^0} \to f$(left) and for $D^0 \to f$(right) produced at the center-of-mass energy of the $\Psi(3770)$.}\label{fig:tagging}
\end{center}
\end{figure}
Where $\Gamma_{+}$ refers to $D^0 (q_c = +2/3)$ decays and $\Gamma_{-}$ to $\overline{D^0} (q_c = -2/3)$ decays, $h_{\pm} =  1 \pm e^{\Delta\Gamma \Delta t}$ and $\lambda_{f} = \frac{q}{p} \frac{\overline{A}}{A}$. 
Here $q$ and $p$ are the parameters defining the mixing and $A$ ($\overline{A}$) is the amplitude for the $D$ ($\overline{D}$) decay to a final state $f$. If $|A|^2 \neq |\overline{A}|^2$ there is direct $CP$ violation (in the decay) and $|q/p|\neq 1$ would signify $CP$ violation in mixing. The study of $\lambda_{f}$ (which should not be confused with the term $\lambda$ appearing in the CKM matrix) is able to probe the combination of $CP$ violation due to mixing and decay, and this form of $CP$ violation is referred to as $CP$ violation in the interference between mixing and decay. Considering Eqns.~(\ref{EQ:p0toffinal},\ref{EQ:p0toffinaldeltat}) the time-dependent asymmetries associated with the time evolution of the $D^0$ mesons can be written in terms of the physical decay rate including the mistag probability, $\omega (\bar\omega)$, for incorrect tagging of the \Dz (\Dzb) flavour as follows
\begin{eqnarray}
\Gamma^{Phys}(t) &=& (1-\omega)\Gamma_{+} (t) + \overline{\omega}\, \Gamma_{-}(t),\\
\overline{\Gamma}^{Phys}(t) &=& \omega\Gamma_{+} (t) + (1 - \overline{\omega})\Gamma_{-}(t),
\end{eqnarray}
where $\Gamma_{+}(t)$ and $\Gamma_{-}(t)$ are from Eqns.~(\ref{EQ:p0toffinal}) and~(\ref{EQ:p0toffinaldeltat}) and $\omega$ ($\overline\omega$) represents the mistag probability for the particle (anti-particle) apparent decay rates for \Dz and \Dzb, respectively. Hence for un-correlated mesons the time dependent $CP$ asymmetry accounting for mistag probability is
\begin{widetext}
\begin{eqnarray}
{\cal A}^{Phys}(t) &=& \frac{\overline{\Gamma}^{Phys}(t) - \Gamma^{Phys}(t) } { \overline{\Gamma}^{Phys}(t) + \Gamma^{Phys}(t)}=\Delta \omega + \frac{ (D - \Delta\omega)e^{\Delta \Gamma t/2}[ (|\lambda_{f}|^2 - 1)\cos\Delta M t + 2 Im\lambda_{f} \sin\Delta M t ]}{h_+ (1+|\lambda_{f}|)^2/2 + Re(\lambda_{f}) h_-},
  \label{eq:asymtagging}
\end{eqnarray}
\end{widetext}
where $\Delta\omega =\omega - \overline\omega$ and $D=1-2\omega$.

Similarly the asymmetry for correlated mesons is
\begin{widetext}
\begin{eqnarray}
{\cal A}^{Phys}(\Delta t) = \frac{\overline{\Gamma}^{Phys}(\Delta t) - \Gamma^{Phys}(\Delta t) } { \overline{\Gamma}^{Phys}(\Delta t) + \Gamma^{Phys}(\Delta t)}= -\Delta \omega + \frac{ (D + \Delta\omega)e^{\Delta \Gamma \Delta t/2}[ (|\lambda_{f}|^2 - 1)\cos\Delta M\Delta t + 2 Im\lambda_{f} \sin\Delta M \Delta t ]}{h_+ (1+|\lambda_{f}|)^2/2 + Re(\lambda_{f}) h_-}.
  \label{eq:asymtagging1}
\end{eqnarray}
\end{widetext}

The above equations may be written in terms of $x$ and $y$ allowing for the measurement of the mixing phase. We report here the time-dependent asymmetry equation for correlated mesons (similar results may be obtained in the un-correlated case):
\begin{widetext}
\begin{eqnarray}
{\cal A}^{Phys}_{x,y}(\Delta t) = -\Delta \omega + \frac{ (D + \Delta\omega)e^{y\Gamma \Delta t}[ (|\lambda_{f}|^2 - 1)\cos x\Gamma\Delta t + 2 Im\lambda_{f} \sin x\Gamma \Delta t ]}{h_+ (1+|\lambda_{f}|)^2/2 + Re(\lambda_{f}) h_-}.
  \label{eq:asymtaggingxy}
\end{eqnarray}
\end{widetext}

\section{MC test of Time-dependent numerical analysis}
\label{sec:tdna}
One of the issues raised in~\cite{bim} is the possibility to use different decay channels of the $D^0$ mesons to constrain the value of the angle $\beta_c$ of the $charm$ triangle. The decay $D^0 \to K^+ K^-$ will be used to measure the mixing phase, the decay $D^0 \to \pi^+ \pi^-$ will be used to measure $\phi_{MIX}-2\beta_c$ and the difference between the two channels will provide a first measurement of the angle $\beta_c$. In this framework, long distance contributions to decay are not considered. The latter together with the different contribution to decay $D^0 \to \pi^+\pi^-$ from $penguin$ topologies will introduce theoretical uncertainties, and for this reason we refer to the angle $\beta_c$ as $\beta_{c,eff}$ where $effective$ indicates that there are theoretical uncertainties that need to be evaluated. To evaluate the asymmetry, and estimate the precision on $\beta_{c,eff}$ that one might achieve in the different experimental environments described in the previous section, we generate a set of one hundred Monte Carlo data samples. Each one based on the expected number of tagged $D^0$ decays in the corresponding experimental setup, and we generate data according to the distributions given in Eqns.~(\ref{EQ:p0toffinal}) and (\ref{EQ:p0toffinaldeltat}), where the parameters involved are evaluated as in  Ref.~\cite{pdg}. We evaluate the asymmetry given in Eqns.~(\ref{eq:asymtagging}) and~(\ref{eq:asymtagging1}) including the expected mistag probabilities, and perform a binned fit to the simulated data. The distributions that we are considering here have been expressed as functions of $|\lambda_{f}|$ and $arg(\lambda_f) \equiv\phi=\phi_{MIX}-2\phi_{CP}$, and the fit is performed keeping $|\lambda_{f}|=1$ and allowing $arg(\lambda_f)$ to vary. The same results are obtained when also $|\lambda_{f}|$ is also allowed to vary in the fit. It is important to mention that a measurement of $\lambda_{f} \neq 1$ in an experiment would be a signature of $CP$ violation~\cite{bim}.

\subsection{Super$B$ at the $\Upsilon(4S)$}
\label{sec:superb4s}
\begin{figure}[!ht]
\begin{center}
\resizebox{8.cm}{!}{
\includegraphics{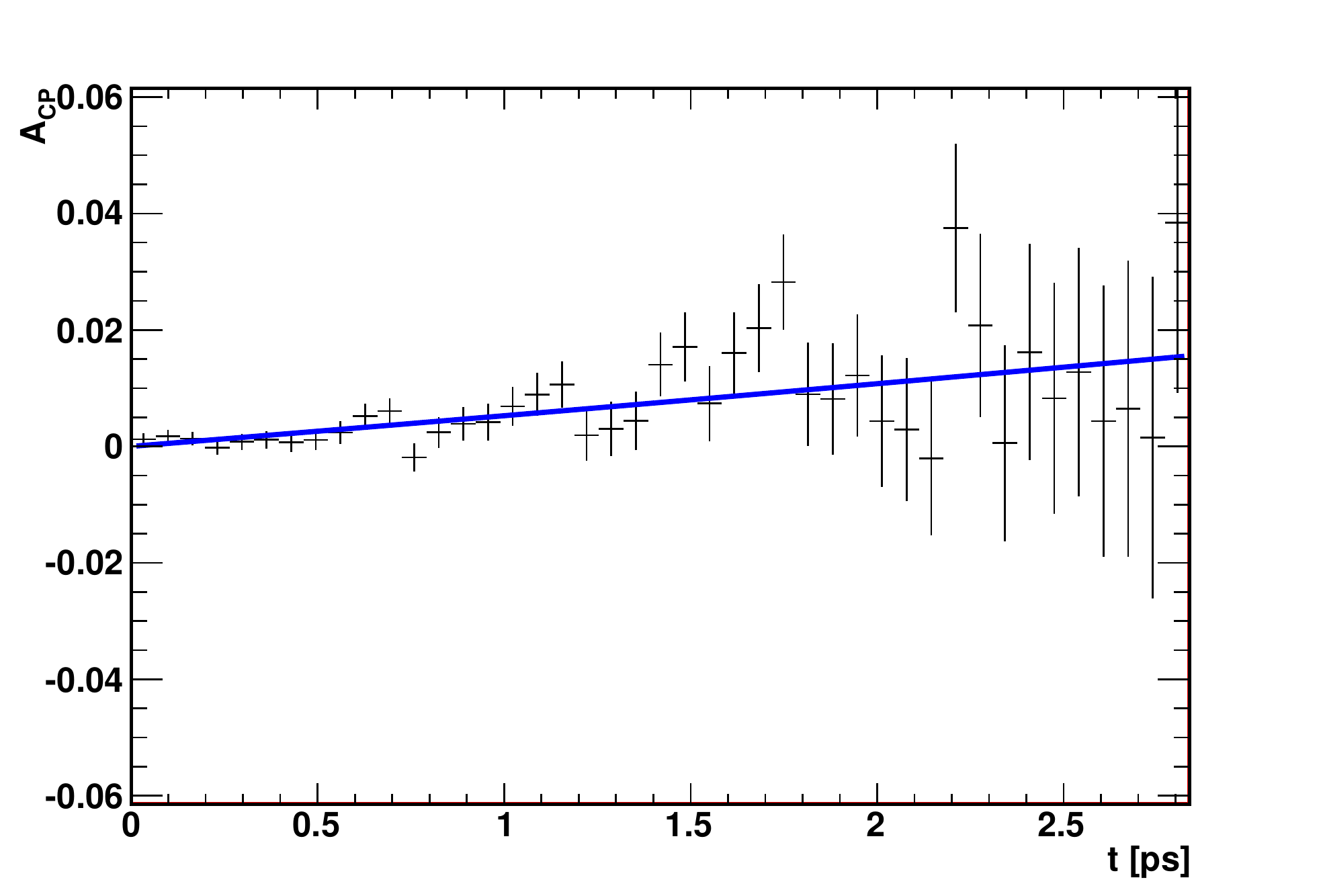}
}
\caption{The time-dependent $CP$ asymmetry expected for $D^0 \to \pi^+ \pi^-$ decays in a $75 \invab$ sample of data at the $\Upsilon(4S)$.}\label{fig:4S}
\end{center}
\end{figure}

The Super$B$ collaboration is expected to start taking data in 2017 ~\cite{superb1}~\cite{superb2}~\cite{superb3}~\cite{superb4}, and the integrated luminosity which will be achieved with the full program is expected to be $75 \invab$. With this luminosity one would expect to reconstruct $6.6 \times 10^6$ tagged $D^0\to \pi^+\pi^-$ events
in a data sample of 75\invab with a purity of 98\%~\cite{bim}. The results of the numerical analysis are shown in Fig.~\ref{fig:4S}.  

The asymmetry parameters determined here have a precision of $\sigma_{arg(\lambda_{\pi\pi})}=\sigma_{\phi_{\pi\pi}}=2.2^o$. The same procedure when applied to the $D^0 \to K^+ K^-$ channel to measure $\sigma_{arg(\lambda_{KK})}=\sigma_{\phi_{KK}}$, for which one would expect to reconstruct $~1.8\times 10^7$ events, leads to precision of $\sigma_{\phi_{KK}}=1.6^o$. When the results from $D^0 \to K^+ K^-$ and $D^0 \to \pi^+ \pi^-$ are combined one obtains a precision in $\beta_{c,eff}$ of $\sigma_{\beta_{c,eff}}=1.4^o$.

\subsection{Super$B$ at the $\Psi(3770)$}
\label{sec:3770}
\begin{figure}[!ht]
\begin{center}
\resizebox{8.cm}{!}{
\includegraphics{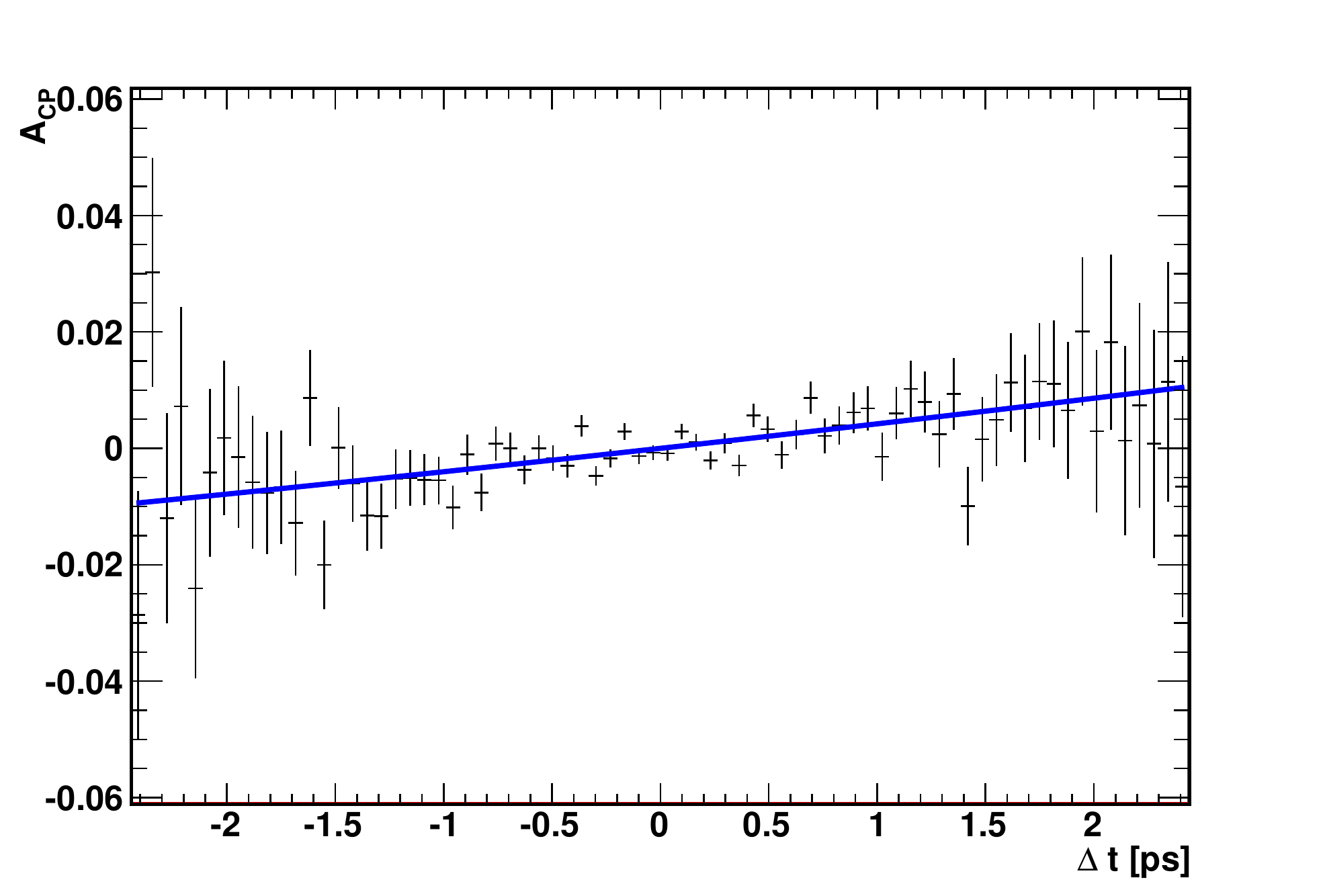}
}
\caption{The time-dependent $CP$ asymmetry expected for $D^0 \to \pi^+ \pi^-$ decays in $1 \invab$ sample of data at the $\Psi(3770)$.}\label{fig:3770}
\end{center}
\end{figure}

The Super$B$ collaboration is planning to have a dedicated run at the center of mass energy of the $\Psi(3770)$ resonance, to collect an integrated luminosity of  $1.0 \invab$. With this luminosity one would expect to record ~979000 $D^0\to \pi^+\pi^-$ reconstructed events, when the full set of semi-leptonic decays $K^{(*)} \ell \nu_\ell$ $\ell = e, \mu$ is used to tag the flavor of $D^0$ mesons (with negligible mistag probability). The results of the numerical analysis are shown in Fig.~\ref{fig:3770}.  

The phase $\phi_{\pi\pi}$  could be measured with a precision of $\sigma_{\phi_{\pi\pi}}=5.7^o$. One may also consider
using hadronically tagged events, for example $D^0\to K^- X$ ($K^+ X$), where $X$ is anything, which corresponds to 54\% (3\%) of all $D^0$ meson decays from which one would expect $\omega \simeq 0.03$, and that the asymmetry in
particle identification of $K^+$ and $K^-$ in the detector will naturally lead to 
a small, but non-zero value of $\Delta \omega$.  We expect that there would be 
approximately 4.8 million kaon tagged $D^0\to \pi^+\pi^-$ events in 1.0\invab 
at charm threshold.  Using these data alone, one would be able to measure
$\phi_{\pi\pi}$ to a precision of $2.7^\circ$. Hence if one combines the results from 
semi-leptonic and kaon tagged events, a precision of $\sigma_{\phi_{\pi\pi}}\sim 2.4^\circ$
is achievable.

\subsection{LHCb}
\label{sec:LHCB}
\begin{figure}[!ht]
\begin{center}
\resizebox{8.cm}{!}{
\includegraphics{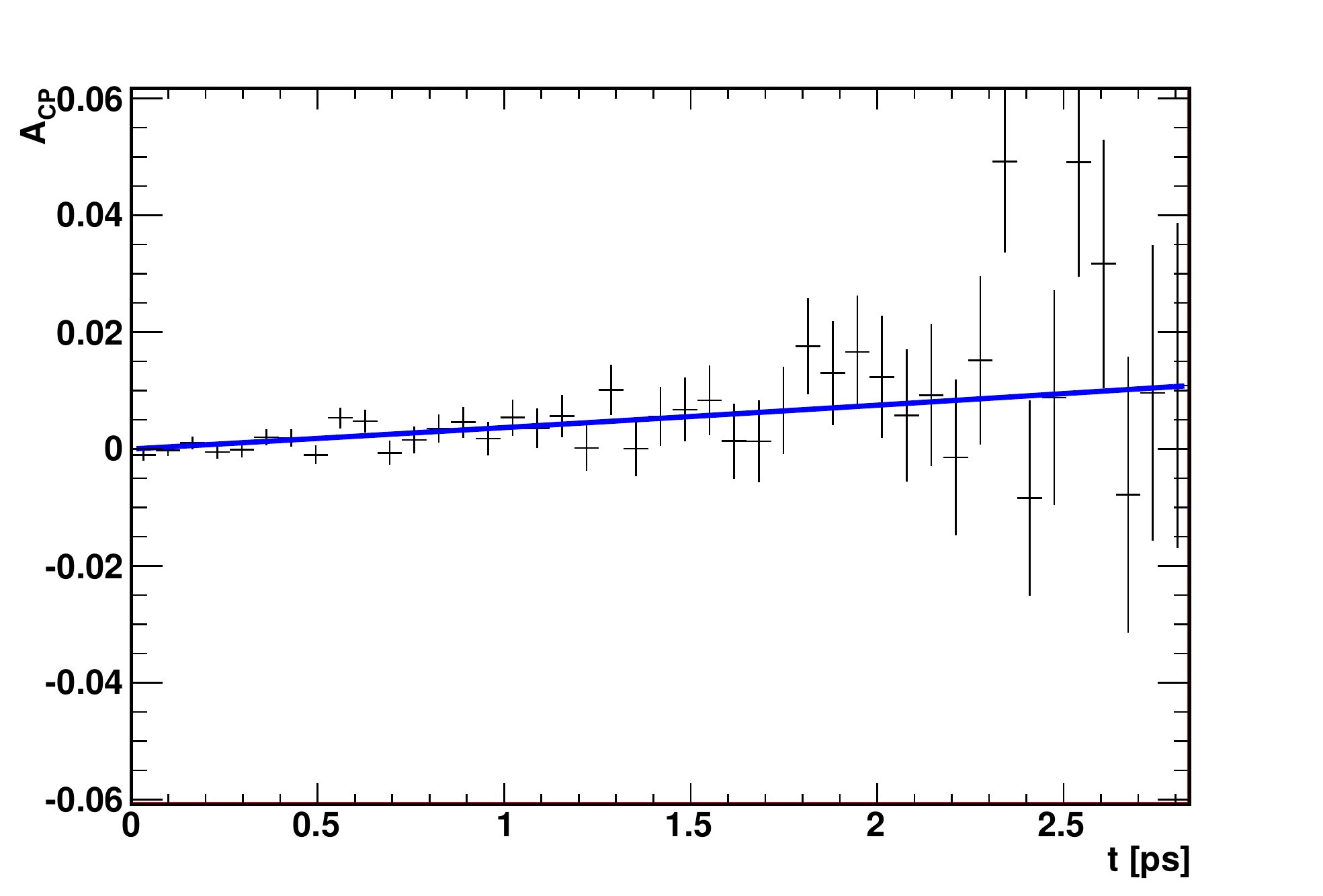}
}
\caption{The time-dependent $CP$ asymmetry expected for $D^0 \to \pi^+ \pi^-$ decays in a $5\invfb$ sample of data at LCH$b$.}\label{fig:LHC}
\end{center}
\end{figure}

Another possible scenario is that of measuring time-dependent asymmetries from uncorrelated
$D$ mesons in a hadronic environment, in particular the LHCb experiment. Here dilution and background effects 
will be larger than those at an $e^+e^-$ machine, but the data are already available and it would be interesting to perform the time-dependent analysis, especially after the recent results on time integrated $CP$ violation in Ref.~\cite{note}. As already mentioned a measurement of $|\lambda_{f}| \neq 1$ will signify $CP$ violation. Given that the measurement of $\lambda_{f}$ is likely  
expected to be dominated by uncertainties, especially in $\omega$ and $\Delta\omega$, it is not clear what the ultimate precision obtained 
from LHCb will be.  The best way to ascertain this would be to perform the measurement on the existing data set.  
We have estimated that LHCb will collect 
$4.9 \times 10^6$ $D^*$ tagged $D^0 \to \pi^+\pi^-$ decays in 5\invfb of data, based on the 
0.62\invfb of data shown in~\cite{note}, and we consider also the outcome of a measurement for 1.1\invfb (equivalent to $0.7 \times 10^6$ $D^*$ tagged $D^0 \to \pi^+\pi^-$ decays) already available after the 2011 LHC run. In~\cite{bim} we estimate a purity of $\simeq 90\%$ and
$\omega \simeq 6\%$ which results in the asymmetry obtained in Fig.~\ref{fig:LHC} for 5\invfb of data.

This fit is translated into a potential measurement of the phase $\phi_{\pi\pi}$ with a precision of $3.0^\circ$. With 1.1\invfb of data we estimate that LHCb may be able to reach a precision of $8^\circ$ on $\phi_{\pi\pi}$.

\subsection{Belle II}
\label{sec:belleII}
\begin{figure}[!ht]
\begin{center}
\resizebox{8.cm}{!}{
\includegraphics{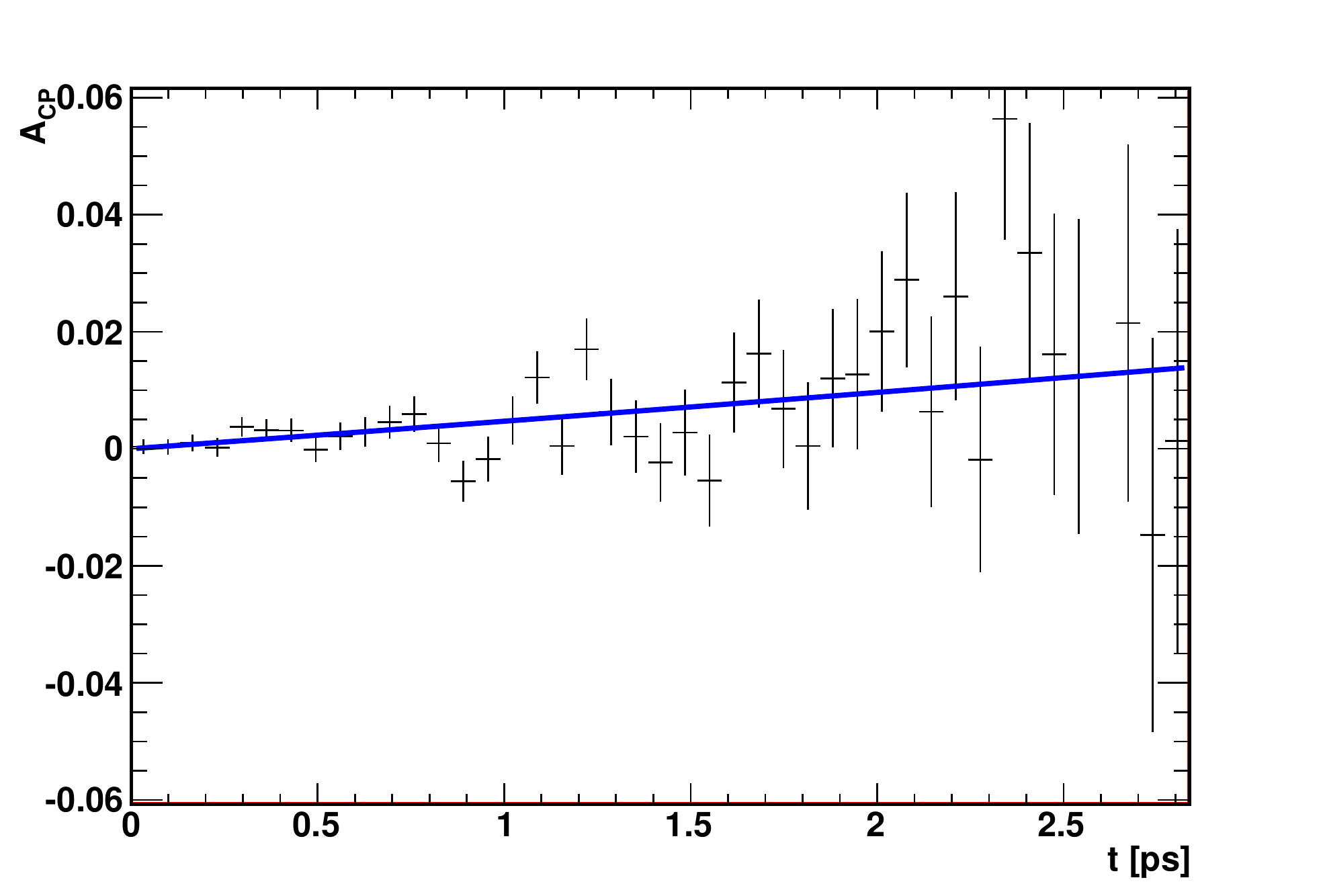}
}
\caption{The time-dependent $CP$ asymmetry expected for $D^0 \to \pi^+ \pi^-$ decays in a $50\invab$ sample of data at the $\Upsilon(4S)$ at Belle II.}\label{fig:belle}
\end{center}
\end{figure}

The last scenario considered here is that of Belle II with 50\invab of data collected at the center of mass energy of the $\Upsilon(4S)$~\cite{belle}. We have considered the same efficiency and mistag probability as for Super$B$ and we expect that $4.4 \times 10^6$ $D^*$ tagged $D^0 \to \pi^+\pi^-$ will be collected. The resulting asymmetry is shown in Fig.~\ref{fig:belle}.
The precision on $\phi_{\pi\pi}$ obtained for this scenario is estimated to be $2.8^\circ$.

\section{Time-dependent sensitivity studies}
\subsection{Sensitivity to $x$}
\label{sec:xpar}
We consider the same data sample discussed in the previous sections for $D^0 \to \pi^+ \pi^-$ and $D^0 \to K^+ K^-$. While we find that results from the time-dependent analysis are not sensitive to the parameter $y$, and that with 1.0 \invab of data collected at charm threshold at Super$B$ it will be possible to improve the currently known precision on $x$ by a factor of two with respect to the most recent HFAG values~\cite{hfag}. The precision that could be reached is shown in Table~\ref{tbl:deltax}.
\begin{table}[!ht]
\caption{Estimates of the sensitivity on $x$ for all the experimental scenarios and their projected luminosities for the decays $D^0 \to \pi^+ \pi^-$ and $D^0 \to K^+ K^-$ and $\phi=\phi_{MIX}-2\beta_{c,eff}$.}\label{tbl:deltax}

\begin{center}
    \begin{tabular}{ | l | c | c |}
    \hline
    Experiment/HFAG & $\sigma_x (\phi = \pm 10^o)$ & $\sigma_x (\phi = \pm 20^o)$  \\ \hline \hline
    Super$B$ [$\Upsilon(4S)$]  &  & \\ 
    $D^0 \to \pi^+ \pi^-$ & $0.12\%$& $0.06\%$      \\  
    $D^0 \to   K^+   K^-$ & $0.08\%$& $0.04\%$      \\ \hline \hline
    Super$B$ [$\Psi(3770)$]  &  & \\ 
    $D^0 \to \pi^+ \pi^- (SL)$ & $0.30\%$& $0.15\%$    \\  
    $D^0 \to \pi^+ \pi^- (SL+K)$ & $0.13\%$& $0.06\%$    \\ 
    $D^0 \to   K^+   K^- (SL)$ & $0.19\%$& $0.10\%$    \\ 
    $D^0 \to   K^+   K^- (SL+K)$ & $0.08\%$& $0.04\%$    \\ \hline \hline
    LHCb &  & \\ 
    $D^0 \to \pi^+ \pi^-$ (1.1 \invfb)& $0.40\%$& $0.20\%$    \\  
    $D^0 \to   K^+   K^-$ (1.1 \invfb)& $0.22\%$& $0.11\%$    \\
    $D^0 \to \pi^+ \pi^-$ (5.0 \invfb)& $0.15\%$& $0.08\%$    \\  
    $D^0 \to   K^+   K^-$ (5.0 \invfb)& $0.09\%$& $0.04\%$    \\ \hline \hline
    Belle II & & \\
    $D^0 \to \pi^+ \pi^-$ & $0.14\%$ & $0.07\%$   \\  
    $D^0 \to   K^+   K^-$ & $0.10\%$ & $0.04\%$   \\ \hline \hline 
    HFAG & \multicolumn{2}{|c|}{$0.20\%$} \\
  \hline
    \end{tabular}
\end{center}
\end{table}

\subsection{Sensitivity to $\beta_{c,eff}$, $\phi_{MIX}$ and $\phi_{CP}$ }
We show here a summary of the possible sensitivities that the different experiments could achieve when measuring the mixing and the weak phase.

\begin{table}[!ht]
\caption{Summary of expected uncertainties from 1\invab of data at charm threshold, 75\invab of 
data at the \FourS, 5\invfb of data from \lhcb, and 50\invab of data at the \FourS at Belle II for $D^0\to\pi^+\pi^-$ decays.
The column marked SL corresponds to semi-leptonic tagged events, and 
the column SL+K corresponds to semi-leptonic and kaon tagged events at charm threshold. The last row shows the precision in $\beta_{c,eff}$ expected from a simultaneous fit to $\pi\pi$ and $KK$ where we assume that, for $KK$, the decay is dominated by a tree amplitude.}
\label{tbl:numericalanalysis}

\begin{center}
    \begin{tabular}{ | l | c  c  c |c|c|}
    \hline
    \multirow{3}*{Parameter} &              &     Super$B$      &  & LHCb & Belle II\\
			     &      $\Psi(3770)$        &     $\Psi(3770)$      & $\Upsilon(4S)$ &  & \\
                             &         SL                 &             SL+K &        $\pi^{\pm}_s$        &   $\pi^{\pm}_s$   &     $\pi^{\pm}_s$     \\ \hline \hline
    $\sigma_{\phi_{\pi\pi}}=\sigma_{arg(\lambda_{\pi\pi})}$ & $5.7^\circ$  & $2.4^\circ$ & $2.2^\circ$ & $3.0^\circ$ & $2.8^\circ$  \\  
    $\sigma_{\phi_{KK}}=\sigma_{arg(\lambda_{KK})}$ & $3.5^\circ$ & $1.4^\circ$ & $1.6^\circ$ & $1.8^\circ$ & $1.8^\circ$\\ 
    $\sigma_{\beta_{c,eff}}$ & $3.3^\circ$ & $1.4^\circ$ & $1.4^\circ$ & $1.9^\circ$& $1.7^\circ$\\
    \hline
    \end{tabular}
\end{center}
\end{table}

At first order the decay $D^0 \to K^+K^-$ measure the mixing phase, therefore one can consider $\phi_{KK}=arg(\lambda_{KK})= \phi_{MIX}$ and use the time dependent analysis to measure it to a precision of $\approx 1.4^o-1.6^o$. 

\subsection{Systematic uncertainties}
The knowledge of the parameters $x$ and $y$ which define the mixing is limited by their relative uncertainties. Since our analysis is not sensitive to the parameter $y$, we considered the most recent results from the HFAG~\cite{hfag} and we evaluated the effect of varying the parameter $\Delta\Gamma= 2y\Gamma$ considering plus-and-minus one standard deviation. This is the systematic uncertainty due to the limited precision in $y$. The value of the uncertainty in the parameter $y$ is 0.013\% and it is given in~\cite{hfag}. The results are shown in Table~\ref{tbl:sys}
\begin{table}[!ht]
\caption{Summary of expected systematic uncertainties due to the limited knowledge of the parameter $y$ from 1\invab of data at charm threshold and 75\invab of 
data at the \FourS. The column marked SL corresponds to semi-leptonic tagged events, and 
the column SL+K corresponds to semi-leptonic and kaon tagged events at charm threshold while $\pi^{\pm}_s$ refers to the slow pion tag at the $\Upsilon(4S)$.}
\label{tbl:sys}

\begin{center}
    \begin{tabular}{ | l | c  c  c |c|c|}
    \hline
    \multirow{2}*{Parameter} &      $\Psi(3770)$        &     $\Psi(3770)$      & $\Upsilon(4S)$ \\
                             &         SL                 &             SL+K &        $\pi^{\pm}_s$  \\ \hline \hline
    $\sigma_{\phi_{\pi\pi}} (sys.)$ & $0.5^\circ$  & $0.2^\circ$ & $0.05^\circ$  \\  
    $\sigma_{\phi_{KK}} (sys.)$  & $0.2^\circ$ & $0.1^\circ$ & $0.02^\circ$\\ 
    $\sigma_{\beta_{c,eff}} (sys.) $ & $0.27^\circ$ & $0.11^\circ$ & $0.03^\circ$\\
    \hline
    \end{tabular}
\end{center}
\end{table}

\subsection{Combined results for Super$B$}
We evaluated the combination of the results obtained for the different centre of mass energy at Super$B$. The final results are made on the assumption that $\phi=\phi_{MIX}-2\beta_c=\pm 10^o$ and they are shown in Table~\ref{tbl:final}

\begin{table}[!ht]
\caption{Combined sensitivities at Super$B$.}\label{tbl:final}

\begin{center}
    \begin{tabular}{ | l | c | c |}
    \hline
\multirow{2}*{Parameter} &      Statistical        &  Systematic   \\ 
                         & sensitivity             & sensitivity  \\ \hline \hline
        $\sigma_x$ ($D^0 \to \pi^+\pi^-$)& $0.09\%$ & -  \\
	$\sigma_x$ ($D^0 \to K^+K^-$)&   $0.05 \% $ & -\\ \hline \hline
	$\sigma_{\phi_{\pi\pi}}$ &  $1.62^o $ & $0.14^o$\\
	$\sigma_{\phi_{KK}}$ &   $1.05^o$ & $0.02^o$ \\
	$\sigma_{\beta_{c,eff}}$ &  $0.92^o$ &$0.03^o$ \\
	\hline
    \end{tabular}
\end{center}
\end{table}

\section{Conclusions}
\label{sec:conclusions}
This paper elucidates the time-dependent analysis of the $D^0$ mesons discussed in Ref.~\cite{bim}. We concentrated on the possible measurement of the $\beta_{c,eff}$ angle of the charm unitarity triangle, on the mixing phase $\phi_{MIX}$ and on the mixing parameters. We estimate our results and compare them for the experimental environments that we think could and should perform this analysis: Super$B$, LHCb and Belle II. We found that Super$B$ may perform better this analysis, but time is required before the collaboration will start data taking. LHCb will have to control the background levels to perform this measurement resulting then in a challenging analysis. However as referred to in the article the LHCb collaboration has already available an amount of data to analyse. This same amount of data has already shown a first hint of direct $CP$ violation in charm, we think it would be worth go through the time-dependent formalism. The Belle II collaboration will start data taking in few years, and the background-clean environment will allow to perform a time-dependent analysis and an evaluation of the mixing phase and of the $\beta_{c,eff}$ at high precision.

\section{Acknowledgments}
This work has been supported by Queen Mary, University of London.

\end{document}